\documentclass[sigplan,screen,10pt,dvipsnames,svgnames,x11names,table,tempa,nonacm]{acmart}
\renewcommand\footnotetextcopyrightpermission[1]{} 
\usepackage[T1]{fontenc}
\usepackage{hyperref}
\usepackage{amsmath,amsfonts}
\usepackage{algorithmic}
\usepackage{graphicx}
\usepackage{textcomp}
\usepackage{balance}
\usepackage{xcolor}
\usepackage{epsfig}
\usepackage{colortbl}
\usepackage{url,color}
\usepackage{breakurl}

\usepackage{subfig}
\usepackage{array}
\usepackage{arydshln}
\usepackage{multirow}
\usepackage{makecell}
\usepackage{xspace}
\usepackage{transparent}
\usepackage{verbatim}
\usepackage{wrapfig}
\usepackage[english]{babel}
\usepackage{tikz}
\usepackage{listings}
\usepackage{diagbox}
\usepackage{pifont}
\usepackage[inkscapepath=./figures]{svg}
\usepackage{csvsimple}
\usepackage{hyperref}
\usepackage{cleveref}
\usepackage[abbreviations]{foreign}
\usepackage[pscoord]{eso-pic}
\usepackage{fancyhdr}

\def\BibTeX{{\rm B\kern-.05em{\sc i\kern-.025em b}\kern-.08em
    T\kern-.1667em\lower.7ex\hbox{E}\kern-.125emX}}

\newcommand{\projectname}{\textsc{TEE\-Mon}\xspace}
\newcommand{\sys}{\projectname}
\newcommand{\scone}{\textsc{SCONE}\xspace}
\newcommand{\ignore}[1]{}

\PassOptionsToPackage{usenames,dvipsnames}{xcolor}

\newcommand{\myparagraph}[1]{\smallskip{}\noindent{\bf{}{#1}.\hspace{.2em}}\nopagebreak{}}
\newcommand{\tightparagraph}[1]{\noindent{\bf{}{#1}.\hspace{.2em}}\nopagebreak{}}
\newcommand{\subheading}[1]{\vspace{2mm}\noindent{\bf{}{#1}}\\\noindent{}}

\newcommand{\out}[1] {}

\newcounter{codeLineCntr}

\reversemarginpar
\newif\ifnotes
\notestrue

\newcommand{\punt}[1]{}

\renewcommand{\eqref}[1]{Equation~(\ref{eq:#1})}

\newcommand{\proc}[1]{\ifmmode\mbox{\textsc{#1}}\else\textsc{#1}\fi}

  \newcommand{\func}[1]{\ifmmode\mathrm{#1}\else\textrm{#1}fi} %

\renewcommand{\,}{\hspace{.1em}}

\makeatother
\usepackage{tcolorbox}
\title{\textsc{TEEMon}: A continuous performance monitoring framework for TEEs}

\author{Robert Krahn}
\affiliation{\institution{TU Dresden}}
\author{Donald Dragoti}
\affiliation{\institution{TU Dresden}}
\author{Franz Gregor}
\affiliation{\institution{TU Dresden, Scontain UG}}
\author{Do Le Quoc}
\affiliation{\institution{TU Dresden, Scontain UG}}
\author{Valerio Schiavoni}
\affiliation{\institution{Universit\'e de Neuch\^atel}}
\author{Pascal Felber}
\affiliation{\institution{Universit\'e de Neuch\^atel}}
\author{Clenimar Souza}
\affiliation{\institution{Universidade Federal de Campina Grande}}
\author{Andrey Brito}
\affiliation{\institution{Universidade Federal de Campina Grande}}
\author{Christof Fetzer}
\affiliation{\institution{TU Dresden, Scontain UG}}

\begin{abstract}
Trusted Execution Environments (TEEs), such as Intel Software Guard eXtensions (SGX), are considered as a promising approach to resolve security challenges in clouds.
TEEs protect the confidentiality and integrity of application code and data even against privileged attackers with root and physical access by providing an isolated secure memory area, \ie{}, \emph{enclaves}.
The security guarantees are provided by the CPU, thus even if system software is compromised, the attacker can never access the enclave's content. 
While this approach ensures strong security guarantees for applications, it also introduces a considerable runtime overhead in part by the limited availability of protected memory (enclave page cache).
Currently, only a limited number of performance measurement tools for TEE-based applications exist and none offer performance monitoring and analysis during runtime.

This paper presents \sys{}, the first continuous performance monitoring and analysis tool for TEE-based applications.
\sys{} provides not only fine-grained performance metrics during runtime, but also assists the analysis of identifying causes of performance bottlenecks, \eg{}, excessive system calls.
Our approach smoothly integrates with existing open-source tools (\eg, Prometheus or Grafana) towards a holistic monitoring solution, particularly optimized for systems deployed through Docker containers or Kubernetes and offers several dedicated metrics and visualizations.
Our evaluation shows that \sys{}'s overhead ranges from \(5\)\% to \(17\)\%.

\end{abstract}

\begin{document}
\thispagestyle{plain}
\thispagestyle{fancy}
\maketitle

\section{Introduction}\label{sec:introduction}
Cloud computing is a popular way to deploy modern online services since it lets users focus on their applications by delegating tasks like resource management to the cloud provider.
In such multi-tenant environments, users care about protecting their data and applications, especially under the threats of powerful adversaries with root privileges or even physical access to machines.
As such, users cannot rely on OS access control mechanisms (bypassed by the former), or on process/memory isolation (bypassed by the latter, for instance via cold-boot attacks~\cite{halderman2009lest}).
In this context, cloud users have to ensure their data is always protected: at rest, in transit over the wire, and while being processed in the main memory.

To overcome these issues, hardware-assisted Trusted Execution Environments (TEEs) such as Intel SGX~\cite{Anati2013a,McKeen2013}, ARM TrustZone~\cite{TrustZone}, IBM SecureBlue++~\cite{SecureBlue1,SecureBlue2}, and AMD SEV~\cite{SEV} offer a practical approach to protect their services in an untrusted cloud~\cite{mofrad2018comparison}.
The TEE technologies provide strong integrity and confidentiality guarantees regardless of the trustworthiness of the underlying software (\eg{}, the operating system or the hypervisor).
Recently, Intel SGX has become available in the cloud~\cite{IBMCloudSGX,DCsv2}, unleashing a plethora of services to be ported, including data processing systems such as MapReduce~\cite{vc3}, coordination services~\cite{brenner2016securekeeper}, content-based routing~\cite{pires2016secure} and databases~\cite{sartakov2018stanlite, enclavedb-a-secure-database-using-sgx}.
Legacy services can also be executed with Intel SGX without any modification by using frameworks~\cite{scone, Tsai:2017} that transparently shield existing applications. 

While promising at first glance, the approach of leveraging TEEs suffers from several technical issues, especially regarding performance overhead.
Indeed, legacy applications running inside secure isolated areas of the hardware, called \emph{enclaves}, suffer from significant performance issues~\cite{Weichbrodt:2018,scone}.
From the cloud user's vantage point, they require strong security guarantees for the code and data of their applications, as well as low runtime overhead.
Gjendrum \etal{}~\cite{Gjerdrum:2017} proposed a set of guidelines for better enclave performance in a cloud environment recommending small enclave page cache (EPC) sizes of \(64\)~kB and minimizing \textit{ECALL} arguments for faster transitions.
Several studies~\cite{scone,Weichbrodt:2018,brenner2016securekeeper} identified costly EPC paging and enclave transitions as major SGX performance bottlenecks.
While solutions exist to mitigate this issue (\eg{}, Switchless~\cite{tian-systex18-switchless}, HotCalls~\cite{Kim-eurosys19-hotcalls}), performing a context switch from the inside to the outside of enclaves still introduces a significant overhead since the hardware needs to prevent the context switch from revealing any sensitive data stored inside enclaves.

To profile the performance of SGX enclaves, Intel provides the VTune Amplifier~\cite{VTune}.
This tool is designed for profiling applications at the instruction level.
While helpful, it lacks performance insights regarding the specific SGX performance metrics.
SGX-Perf~\cite{Weichbrodt:2018} and TEE-Perf~\cite{TEE-Perf} overcome this limitation by allowing developers to trace enclave execution and providing SGX performance metrics such as enclave transitions and paging.
However, SGX-Perf supports only applications that use the Intel-SGX Software Development Kit.
In addition, both SGX-Perf and TEE-Perf do not support runtime monitoring of the performance overhead of an application running inside an enclave.
Thus, they cannot support SGX frameworks to tune system parameters during runtime to improve the performance of the applications, for example by increasing the number of threads on the inside or outside of an enclave.

To summarize, TEEs are widely adopted to provide strong security guarantees for applications running in an untrusted environment, \eg{}, a public cloud.
However, there is a lack of tools to monitor the performance of these applications running inside TEE enclaves at runtime.
We fill this gap by designing and implementing \sys{}, which allows users to monitor the performance of their applications running in TEEs and to identify performance issues and bottlenecks in real-time.
More specifically, \sys{} has the following design features:

\myparagraph{1. Lightweight}
\sys{} incurs low performance overhead while providing useful and accurate performance profiling.

\myparagraph{2. Transparency}
\sys{} does not require modification of the monitored application.

\myparagraph{3. Generality}
\sys{} is framework-agnostic and can be used with many SGX frameworks such as SCONE~\cite{scone}, Gra\-phe\-ne-SGX~\cite{Tsai:2017}, or SGX-LKL~\cite{sgx-lkl}.

We implemented \sys{} to monitor the performance of applications running inside Intel SGX enclaves during runtime.
In this work, we focus on Intel SGX since it is widely used in practice, although the technique can be applied to other TEEs as well.
We evaluated \sys{} using real-world applications and state-of-the-art SGX frameworks.
Our evaluation shows that \sys{} has a negligible overhead  from \(5\)\% to \(17\)\% depending on running applications while providing useful performance metrics data for users with an intuitive visualization during runtime.
In addition, to the best of our knowledge, this work is the first providing an intensive performance comparison between state-of-the-art Intel SGX frameworks.
Lastly, \sys{} has been tested and integrated with Kubernetes to monitor over \(6000\) distributed SGX enclaves in production, where it was used to track various performance metrics and has allowed the developers of the SCONE framework to continuously monitor the impact of different software generations on a detailed level (e.g. consumption of the enclave page cache, page faults, etc)~\cite{teemonscone}.
\vspace{-2mm}\section{Related Work}\label{sec:relatedwork}

There exist only a few tools to profile and monitor applications executed in TEEs at runtime. 
We briefly report on them, as well as more generic profilers and monitoring tools.
Table~\ref{tab:tools} summarizes our survey.


\newcommand{\YES}{\textcolor{Green4}{\ding{51}}}
\newcommand{\NO}{\textcolor{Red4}{\ding{55}}}

\newcommand{\LAMBDA}{\includesvg[width=5pt]{figures/lambda}}
\newcommand{\BELL}{\includesvg[width=5pt]{figures/bell}}
\newcommand{\CUBE}{\includesvg[width=5pt]{figures/cube}}
\newcommand{\GEARS}{\includesvg[width=7pt]{figures/gears}}

\begin{table}[t!]
\scriptsize
\small
\setlength{\tabcolsep}{3pt}
\setlength{\extrarowheight}{1pt}
\centering
\rowcolors{1}{gray!10}{gray!0}
\begin{tabular}    {
    l
    >{\centering\arraybackslash}p{0.12\columnwidth}
    >{\centering\arraybackslash}p{0.1\columnwidth}
    >{\centering\arraybackslash}p{0.12\columnwidth}
    >{\centering\arraybackslash}p{0.13\columnwidth}
    >{\centering\arraybackslash}p{0.09\columnwidth}
    >{\centering\arraybackslash}p{0.1\columnwidth}
    }
  \rowcolor{gray!25}
  \textbf{Tool} & Frame-work Agnostic & Paging & Enclave Transitions & Orches-trated Applications & Real-Time Reports & Granu-larity\\
\toprule{}
    \hspace{-2pt}LIKWID
    & \YES{}
    & \NO{}
    & \NO{}
    & \YES{}
    & \NO{}
    & \LAMBDA{},\GEARS{}\footnotemark{}
    \\
\texttt{perf}
    & \YES{}
    & \NO{}
    & \NO{}
    & \NO{}
    & \NO{}
    & \LAMBDA{},\GEARS{}
    \\	
MemProf
    & \YES{}
    & \NO{}
    & \NO{}
    & \NO{}
    & \NO{}
    & \CUBE{}
    \\
TEE-Perf
    & \YES{}
    & \NO{}
    & \NO{}
    & \NO{}
    & \NO{}
    & \LAMBDA{}
    \\
\texttt{gprof}
    & \YES{}
    & \NO{}
    & \NO{}
    & \NO{}
    & \NO{}
    & \LAMBDA{}
    \\	
VTune
    & \YES{}
    & \NO{}
    & \NO{}
    & \NO{}
    & \NO{}
    & \LAMBDA{}
    \\
SGX-Perf
    & \NO{}
    & \YES{}
    & \YES{}
    & \NO{}
    & \NO{}
    & \BELL{}
    \\
SGXTOP
    & \YES{}
    & \YES{}
    & \YES{}
    & \NO{}
    & \YES{}
    & \BELL{}
    \\
\projectname{}
    & \YES{}
    & \YES{}
    & \YES{}
    & \YES{}
    & \YES{}
    & \LAMBDA{},\BELL{},\GEARS{}
    \\
    \addlinespace[.1em] \toprule
    
\end{tabular}   
\caption{Profile/monitoring tools for SGX. Granularity can be: \LAMBDA=function, \CUBE=object, \BELL=event, \GEARS=system.\vspace{-5pt}}\label{tab:tools}
\end{table}

\footnotetext{System-wide statistics are available in LMS~\cite{Rohl:2017}.}

\vspace{-2mm}\subsection{Profilers}\label{Profiler}

Over the years, numerous profiling tools and approaches have been proposed.
Ball \etal{}~\cite{Ball:1994} first proposed a set of optimal algorithms for program profiling and instruction tracing to reduce the overhead of profilers.
Hardware specific tools such as LIKWID~\cite{Treibig:2010, Rohl:2017} for x86 environments and MemProf~\cite{Lachaize:2012} for NUMA systems make use of hardware counters allowing developers to explore optimization opportunities specific to the underlying system architecture.
Linux \texttt{perf}~\cite{perf} provides low-level system metrics by attaching to tracepoints, performance counters or probes similarly to eBPF~\cite{eBPF}.
But it does not offer support for TEEs and it cannot provide profiling data for applications running inside SGX enclaves.
VTune Amplifier~\cite{VTune}, a commercial analysis tool by Intel, offers in-depth analysis for SGX applications.
With the help of special hardware features, it gives detailed information regarding time spent on each method or function call but it provides no support for SGX specific metrics like EPC paging.
Further, it depends on the Intel hardware architecture and does not support other vendors.
Another similar SGX-specific profiler is SGX-perf~\cite{Weichbrodt:2018}, which provides statistics on enclave entries and exits as well as EPC paging events based on \texttt{kprobes}~\cite{kprobes}.
SGX-Perf is limited to applications using the Intel SDK for SGX and does not support monitoring during runtime since it implemented a two-phased record and report approach.
\sys{} focuses on continuous monitoring during runtime to provide insights into performance helping to identify the root cause of any bottlenecks and revealing opportunities for performance improvement.

The \texttt{gprof}~\cite{gprof} tool can provide developers with an execution profile by counting function invocations as well as the time spent on each function.
However, it is limited to C, Pascal, and Fortran77 programs and it does not offer cross-platform support.
TEE-perf~\cite{TEE-Perf} is an application profiler for trusted execution environments, offering method-level profiling and flame-graph visualizations.
Also, it is platform-independent by using software level counters for performance measurements.
The performance overhead of TEE-perf is up to \(5.7\)\(\times\) higher compared to that of Linux \texttt{perf} as the injected code runs on each function call~\cite{TEE-Perf}.
This, in turn, limits its use to development environments as high overheads are unacceptable in production.
All of the aforementioned profilers offer application-specific performance data, but some are platform-dependent, some language-dependent and some are proprietary.
Besides, most of the profilers mentioned (\ie, TEE-perf, \texttt{gprof}, SGX-Perf, and VTune Amplifier) are adapted for usage in development and testing stages but are not suitable for providing continuous monitoring due to their high-performance overhead.
While developed within a security context, approaches such as SGX-Step~\cite{van2017sgx} could also be used to debug applications in TEEs but typically incur a substantial slowdown.
Note that the state-of-the-art monitoring and performance profilers such as TEE-perf~\cite{TEE-Perf} report an average slowdown factor of 1.9\(\times\)~compared to native SGX executions.
Closely related to \sys{}, SGX-TOP~\cite{sgxtop} continuously displays SGX related metrics, which are also collected by the TEE metric exporter (Section~\ref{sec:TME}).
However, SGX-TOP focuses solely on displaying SGX-related metrics in a terminal, while lacking archival functionality as well as the combination of multiple metrics into one interface as provided by \sys{}.
\sys{} is designed to provide a more holistic monitoring application.
In addtion, SGX-TOP can be used only with single-node applications, whereas \sys{} supports distributed applications deployed in multiple nodes (e.g., using Kubernetes).

\subsection{Distributed Monitoring}\label{Monitoring}

In a distributed computing environment, applications require constant monitoring to be able to recover and fix deployment issues, fix application-specific bugs or to adjust deployment depending on the current load on the system. 
Major cloud providers offer proprietary monitoring solutions to their clients, such as Amazon CloudWatch~\cite{cloudwatch}, Google Cloud's Operations~\cite{googleoperations}, or IBM SysDig~\cite{ibmsysdig,BorelloSysDig}.
These monitoring systems provide hundreds of metrics regarding the cloud infrastructure and applications.
However, those are dependent on the cloud provider and are usually billed depending on the number of metrics requested.
Besides, in multi-cloud infrastructures, one would need to support and maintain multiple monitoring systems depending on the cloud providers. 

As a result, different companies like New Relic~\cite{newrelic}, SolarWinds~\cite{solarwinds}, Datadog~\cite{datadog}, AppDynamics~\cite{appdynamics}, SignalFx~\cite{signalfx} and Dynatrace~\cite{dynatrace} offer managed solutions to monitoring and software analytics for dynamic micro-service architectures.
These options provide a solution to the vendor lock-in issue and offer a comprehensive list of services including application and infrastructure analytics.
They also support ready-made visualizations and machine learning capabilities running on top of accumulated data for better insights, such as SysDig for SGX capable bare-metal servers at IBM-Cloud~\cite{ibmcloud}.
However, none of them were designed to monitor and analyze the performance of applications running inside of TEE enclaves. 
Our proposed framework additionally supports containerized applications and orchestration management system such as Kubernetes (Section~\ref{sec:deployment}) thus, can be integrated into TEE-enabled cloud environments~\cite{vaucher2018sgx}.



\section{Background}\label{sec:background}

\subsection{Intel Software Guard eXtensions (SGX)}\label{subsec:SGX}

SGX is a set of x86 ISA instructions to secure code and data of applications available since the Intel Skylake\footnote{\url{https://software.intel.com/en-us/articles/an-overview-of-the-6th-generation-intel-core-processor-code-named-skylake}} architecture~\cite{McKeen2013, Anati2013a}.
SGX introduces a concept of a secure \emph{enclave}, \ie, a hardware-\-protected memory region for code and data protected by the CPU with confidentiality and integrity features.
A dedicated region of physical memory called the Enclave Page Cache (EPC) is reserved for enclaves.
EPC is protected by an on-chip Memory Encryption Engine (MEE), which transparently encrypts and decrypts the cache lines as they are respectively written to and read from EPC.
SGX provides a call-gate mechanism to control entry into, and exit from, the TEE~\cite{Costan:2016}.
In most processors, the EPC is limited to \(\sim\)\(128\)~MB, and only \(\sim\)\(94\)~MB can be used for applications~\cite{Costan:2016}. 
However, the latest generations of Intel Xeon processors are equipped with a larger sized EPC. Currently, Intel SGX has been offered in clouds~\cite{IBMCloudSGX, AzureSGX}, enabled plenty of confidential cloud-native applications, e.g., analytics systems~\cite{sgx-pyspark, securetf}, key management system~\cite{palaemon}, and secure software update~\cite{TSR}. 

\subsection{SGX Frameworks}


To run legacy applications with Intel SGX without any source code modification, SGX frameworks such as SCONE~\cite{scone}, SGX-LKL~\cite{sgx-lkl}, and Graphene-SGX~\cite{Tsai:2017} can be utilized.

\myparagraph{SCONE and SGX-LKL}\label{SCONE}
SCONE~\cite{scone} is a shielded execution framework using Intel SGX.
To run applications inside SGX enclaves, programs are linked against a modified standard C library, \ie, SCONE \texttt{libc}.
In this model, the whole application is confined to the enclave memory and the interaction with the untrusted system is executed via a system call interface.
SCONE leverages an asynchronous system call mechanism: threads inside of the enclave execute tasks of the application, pushing system calls to the outside of the enclave.
Threads outside of the enclave asynchronously execute the system calls and push results back. 
In addition, SCONE natively integrates with Docker~\cite{docker} to seamlessly deploy micro-service based applications using container images.
A similar approach is implemented by SGX-LKL~\cite{sgx-lkl}, which also provides a framework that links applications against a modified standard C library (\texttt{musl-libc}).
 
\myparagraph{Graphene-SGX}\label{Graphene}
Graphene-SGX~\cite{Tsai:2017} is an open-source SGX implementation of the original Graphene library OS.
Similar to Haven~\cite{haven2015}, it runs a complete library OS inside of SGX enclaves and lets developers run their applications with Intel SGX without any code modifications.
Graphene-SGX facilitates protection through a \texttt{manifest} file that contains user defined security policies and a list of trusted libraries (with their cryptographic SHA-256 hashes) required by the application. 
Thus, the manifest file allows for a fine granular definition of trusted resources.



\subsection{BPF and eBPF}

The scope of the original BSD Packet Filter (BPF)~\cite{McCanne:1993} was limited to network monitoring and packet filtering, by leveraging RISC CPU registers.
It efficiently filters network packets without copying packets to user-space before analyzing them. 
The extended BSD Packet Filter (eBPF) is capable of taking advantage of modern x86 CPU architectures, 64-bit registers, and just-in-time compilation since the Linux kernel 3.15 kernel. 
Although the original BPF project started as a network monitoring tool, nowadays eBPF programs can be attached to a multitude of hooks in the Linux kernel. 
eBPF added methods to load custom programs into the eBPF virtual machine for the kernel.
In addition, it allows for user space programs to interact with \texttt{BPF\_MAP} data structures in the kernel~\cite{ebpfmaps}. 
\texttt{BPF\_MAPS} are generic key/value stores mainly used to share data between kernel and user-space, and between different eBPF programs.

\sys{} uses eBPF to obtain and provide low-level system statistics such as executed system calls or cache misses, as we detail next.

\section{Architecture}\label{sec:design}

\begin{figure}[!t]
	\centering
	\includegraphics[scale=0.35]{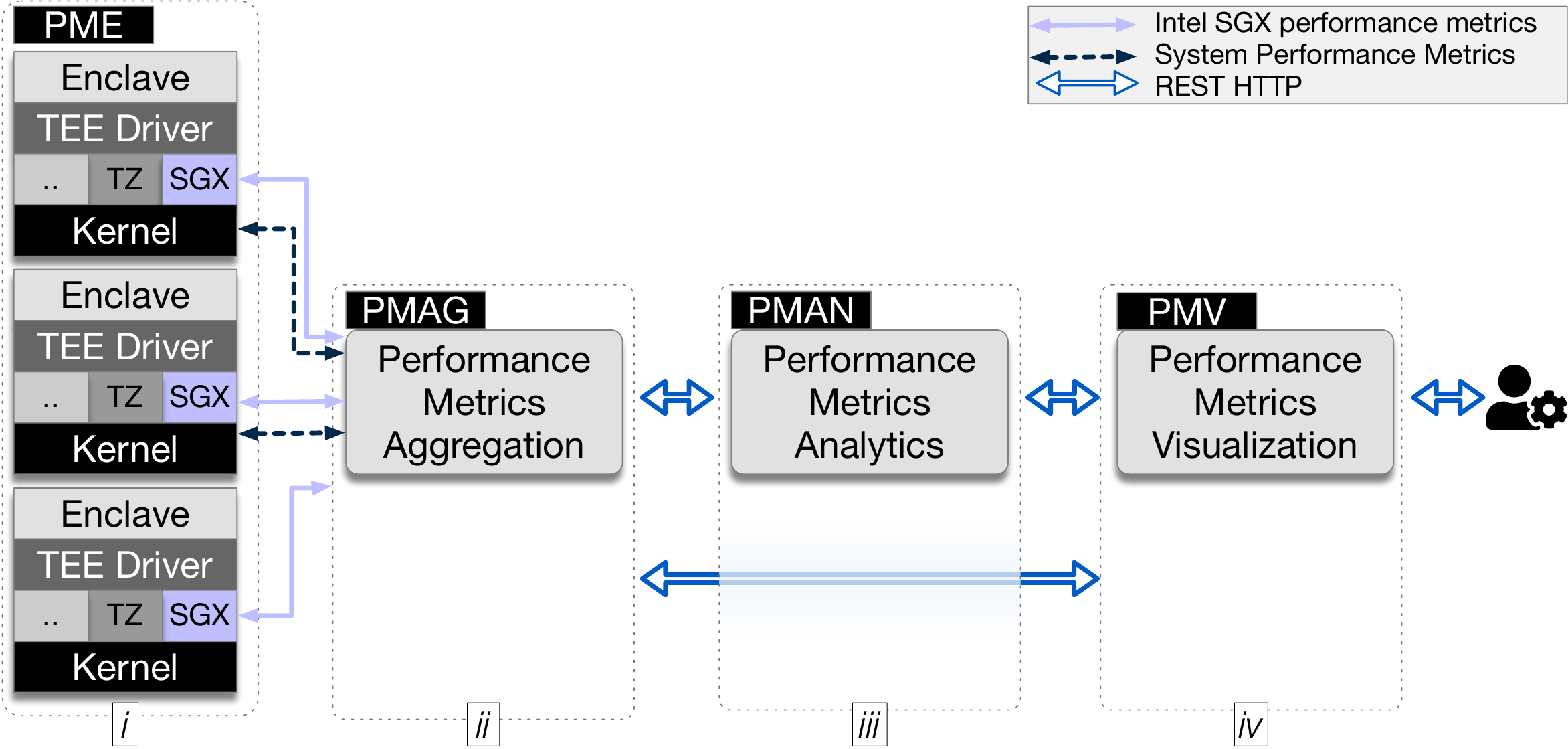}
	\caption{\projectname{} system overview.}\label{fig:system-overview}\label{fig:impl}
	\Description[]{}
\end{figure}
The main goal of \sys{} is to provide a low-overhead monitoring framework able to offer the performance profiler functionality as previous frameworks~\cite{Weichbrodt:2018,TEE-Perf} while allowing users to continuously keep track of their applications running inside SGX enclaves during runtime, a combination of features that as shown earlier is currently not available (Table~\ref{tab:tools}).
In this section, we discuss the architecture of our monitoring framework, explaining the main design decisions.

\sys{} is designed to be generic and applicable to a variety of SGX frameworks without code changes while keeping in mind the state-of-the-art and current best practices in the field of monitoring and observability.
Figure~\ref{fig:system-overview} shows the high level overview of our monitoring framework.

\sys{} consists of four core components: \emph{(i)} The performance metrics exporters (PME) collect metrics at different system levels, including the kernel and the TEE environment (e.g. the Intel SGX driver).
\emph{(ii)} The Performance metrics aggregation (PMAG) combines and stores data in a time series database accessed and processed by \emph{(iii)} the performance metrics analysis (PMAN) component; and \emph{(iv)} the visualization (PMV) component that presents the monitoring data via a web-service to the user.
We provide further details about currently utilized software components in Section~\ref{sec:implementation}. 

In a nutshell, to capture performance metrics with minimal overhead and without code modification of applications, we design an exporter module for the TEE driver to extract the necessary information for keeping track of applications during runtime.
The metrics exporter obtains performance metrics directly from the kernel.
Thereafter, the metrics aggregator combines the monitoring metrics from all distributed exporters to provide a global view of the performance of the monitored applications.

\sys{} uses multiple exporters per host machine with individual tasks, \eg{}, providing SGX-specific or machine-specific metrics.
The analysis component examines the aggregated performance metrics to identify the bottlenecks of monitored applications.
Lastly, the visualization component continuously presents the performance bottlenecks and critical performance metrics in an intuitive way for users via a web-based interface.




\myparagraph{\emph{(i)} Performance Metrics Exporter (PME)}\\[1pt]\noindent
This component combines two modules: the TEE Metrics Exporter which collects TEE related metrics, and the System Metrics Exporter which collects system metrics.
PME exports these collected metrics to the aggregation component in a standardized format.
We describe both next.

\begin{figure}[!t]
	\centering
	\includegraphics[width=0.36\textwidth]{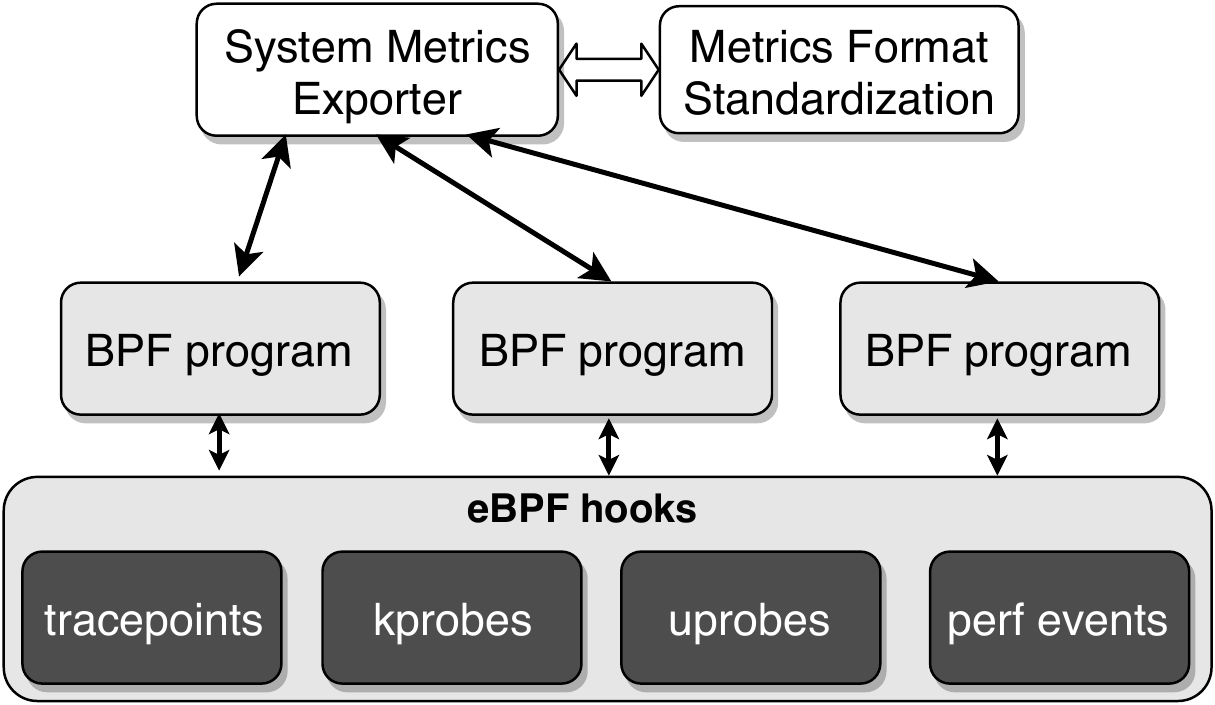}
	\caption{System Metrics Exporter architecture.}\label{fig:SME}
	\Description[]{}
\end{figure}

\vspace{4pt}
\textbf{TEE Metrics Exporter (TME).}\label{sec:TME}
The TME supports two main classes of TEE-related metrics: enclave metrics (initialized enclaves, active enclaves, removed enclaves) and EPC metrics (total EPC pages, free EPC pages, pages marked as old, pages evicted to main memory, pages added to enclaves, pages reclaimed from main memory).
Additional metrics could be added through code-adaptation of the utilized TEE driver.

To capture these TEE metrics with low overhead, the TME connects to specific hooks (\eg{}, \texttt{sgx\_nr\_free\_pages},\\ \texttt{sgx\_nr\_enclaves}, or \texttt{sgx\_nr\_evicted}) in the TEE driver and extracts the data regarding the trusted enclave regions and other specific information to the underlying TEE architecture.
There is a single TME instance per machine that requires privileged access to the underlying operating system. 

While modifying an existing TEE driver would be feasible~\cite{vaucher2018sgx, sgxdriver}, developing an additional component provides a modular architecture in order to support current and future TEE technologies, and it avoids excessive and expensive modifications to the underlying TEE subsystem, as this would need additional effort for maintenance.
In addition, it can also be used to expose additional metrics regarding the platforms running on top of a TEE.
For instance, in Intel SGX, \sys{} can be interchangeably used for SCONE or Graphene-SGX without changing the code of the SGX frameworks.

Our design allows for the PME to be easily customized and used on different TEE platforms as well as for kernel-integrated approaches, such as IBM PEF~\cite{ibmpef}, AMD SEV~\cite{SEV}, or Intel TDX~\cite{inteltdx}.
For these virtual machine based security mechanisms, we envision an extension to the hypervisor, e.g. qemu, that integrates the functionality of the TME.
The extension would, similar to the TME for SGX, export metrics such as the amount of protective memory requested by each virtual machine.
Additionally, other exporters, e.g., the system metrics exporter (SME) described in the next section, can be added to the virtual machine image to export metrics from the guest OS to \sys{}.

\vspace{4pt}
\textbf{System Metrics Exporter (SME).}
The task of the SME (Figure~\ref{fig:SME}) is to collect and export performance metrics from the underlying system infrastructure. 
To obtain low-level system metrics, one needs to access CPU performance counters, kprobes and tracepoints similarly to Linux's \texttt{perf}~\cite{perf}.
This requires writing and running code in kernel space while ensuring it neither corrupts security nor reduces performance. 

To overcome these problems, we use eBPF, the in-kernel virtual machine, allowing kernel instrumentation programs to run in a secure and restricted environment.
The eBPF programs connect to specified hooks in the host's kernel. 
By attaching small eBPF programs to each kernel hook, we can read and extract low-level system statistics and export them to user space via \texttt{BPF\_MAPs}.


The metrics are translated into a standard format understood by the metrics aggregation component (\eg{}, Pro\-me\-the\-us~\cite{prometheus}), and published to its metric endpoint (\eg{}, a web server) to be scraped.
The current \sys{} implementation instruments system calls, context switches, page faults, and last-level cache metrics. 
The SME list of metrics and the corresponding system hooks are shown in Table~\ref{tab:metrics}.
The listed metrics simply provide a guideline on the most important system metrics. 
It can be extended depending on the application's and user's requirements.


\begin{table}[t!]
\small
\setlength{\tabcolsep}{3pt}
\renewcommand\theadfont{\bfseries}
\centering
\begin{tabular}{c;{2pt/1pt}c;{2pt/1pt}c}
  \rowcolor{gray!25}
  \textbf{Type} & \textbf{Method} & \textbf{Field} \\

  \hline
  \thead{Sys. call \\[-2pt] metrics} &
  \makecell{Kernel \\[-2pt] tracepoints} &
  \makecell{\texttt{raw\_syscalls:sys\_enter} \\[-2pt] \texttt{raw\_syscalls:sys\_exit}} \\

  \hline
  \multirow{3}{*}{\thead{Cache \\[-2pt] metrics}} &
  \makecell{Kprobes} &
  \makecell{\texttt{add\_to\_page\_cache\_lru} \\[-2pt] \texttt{mark\_page\_accessed} \\[-2pt] \texttt{account\_page\_dirtied} \\[-2pt] \texttt{mark\_buffer\_dirty}} \\

  \cdashline{2-3}[.5pt/1pt] &
  \makecell{Perf. events} &
  \makecell{\texttt{PERF\_COUNT\_HW\_CACHE\_MISSES} \\[-2pt] \texttt{PERF\_COUNT\_HW\_CACHE\_REFERENCES}} \\

  \hline
  \multirow{3}{*}{\thead{Context \\[-2pt] switches}} &
  \makecell{Perf. events} &
  \makecell{\texttt{PERF\_COUNT\_SW\_CONTEXT\_SWITCHES}} \\

  \cdashline{2-3}[.5pt/1pt] &
  \makecell{Kernel \\[-2pt] tracepoints} &
  \makecell{\texttt{sched:sched\_switches}} \\

  \hline
  \multirow{3}{*}{\thead{Page \\[-2pt] faults}} &
  \makecell{Perf. events} &
  \makecell{\texttt{PERF\_COUNT\_SW\_PAGE\_FAULTS}} \\

  \cdashline{2-3}[.5pt/1pt] &
  \makecell{Kernel \\[-2pt] tracepoints} &
  \makecell{\texttt{exceptions:page\_fault\_user} \\[-2pt] \texttt{exceptions:page\_fault\_kernel}} \\

  \hline
\end{tabular}
\vspace{5pt}
\caption{System Metrics collected by \sys{}.\vspace{-5pt}}\label{tab:metrics}
\vspace{-5pt}
\end{table}

\vspace{-2pt}
\myparagraph{\emph{(ii)} Performance Metrics Aggregation (PMAG)}\\[1pt]\noindent
The PMAG combines and aggregates performance metrics collected from the exporter component (PME).
It embeds a time-series database, a metrics retrieval component, and an HTTP server.
It collects, processes, and aggregates a large number of metrics, from a dynamically changing list of services, with low overhead on the applications.
Using its time-series database, it stores all metrics data samples locally and groups them into chunks for faster retrieval.
Additionally, it allows for multi-dimensional data with the help of metric labels specified as a set of key-value pairs, \eg{}, system calls with a timestamp, system-call number, and the number of occurrences. 

To support the analysis of performance metrics, 
PMAG supports data queries over specified time ranges and labeled dimensions.
It provides detailed quantitative analysis by selecting and applying aggregation functions to query results.

PMAG can connect to every service exposing a metrics endpoint, and users of \sys{} can easily add their application metrics to it. 
For example, a user can instrument applications to export monitoring metrics in the standard text-based format as specified by the OpenMetrics project~\cite{openmetrics} and expose them to a REST endpoint.
From this endpoint, PMAG can collect the measured metrics.

%
%

\vspace{4pt}\textbf{Push vs. Pull in Monitoring.}
There are two main mechanisms used by centralized monitoring systems to gather new metrics: \emph{(i) Push} --- providing an endpoint where services can push their metrics at the rate they happen or \emph{(ii) Pull} --- the monitoring system pulls the metrics from the metrics endpoints on each service periodically. 

The push-based approach allows services to continuously push metrics in real-time and is mainly used by event-based monitoring services like statsd~\cite{statsd}.
However, event-based monitoring can lead to congestion and overloading of the entire monitoring systems during sudden events bursts.

The pull-based approach requires the metrics exporter component to provide an endpoint where the monitoring service can subsequently scrape the metrics at specified intervals. 
Consequently, the resulting data traffic can be managed easier and it lightens the load on the monitoring service itself, compared to having to ingest hundreds of metrics per second as it might happen in the case of system metrics.
Additionally, centralizing data ingestion by the system eliminates the issue of misbehaving services pushing garbage data which might turn into a DoS attack on the monitoring system.
The downside to the pull approach is the need to know which exporters need to be scraped.
This can be solved by using a service discovery system, \ie, a centralized catalog of the running applications and their respective REST endpoints.
The monitoring service also acts as a health checker and can alert in case the monitoring target is unreachable.
\sys{} relies on a pull approach (from the PMAG component) to collect the monitoring data from the TEE Metrics Exporter and System Metrics Exporter components.

\myparagraph{\emph{(iii)} Performance Metrics Analysis (PMAN)}\\[1pt]\noindent{}
Although the PMAG component offers data queries and aggregation functions, it lacks support for metrics visualization and analytics. 
Thus, we design the PMAN component to analyze the aggregated data from the PMAG component in real-time, to identify the bottlenecks or potential anomalies, and to report them to the visualization component for inspection.
In addition, the PMAN component has the ability to aid the identification of bottlenecks in applications running inside TEE enclaves.
Technically, we make use of threshold-based approaches to detect anomalies in monitoring data.
We identified these thresholds using benchmarking with real-world SGX-based applications.
PMAN analyzes the time-series monitoring data using slide window computations, \eg{}, it processes every minute for the last five minutes of the monitoring data.
In each time window, PMAN not only compares the monitoring data with user-defined thresholds to detect anomalies but also provides a box plot for SGX metrics.
PMAN supports handling anomalies in several ways including alerting, dashboard updating, and logging.
Administrators or developers of SGX-based applications can use this information to discover or identify the root of bottlenecks.
\noindent{}Note that PMAN can be further extended to perform more advanced analytics, such as the correlation between SGX metrics and configuration parameters of applications, or performance prediction.

\myparagraph{\emph{(iv)} Performance Metrics Visualization (PMV)}\\[1pt]\noindent
Visualizations are crucial for monitoring services, especially when dealing with complex time-series metrics to spot underlying issues or to quickly infer performance trends. 
Additionally, in cases of failures or incidents, it is useful to limit the view of data to a specific time frame. 
However, choosing the right type of visual representation is not trivial, as it depends as much on individual preferences as on the metrics that a user is trying to visualize.
The PMV component currently supports several visualization options, \eg{}, graphs, histograms, gauges, gradient fills, tables, \etc.
It is also possible to group different metrics so that metrics from the same service or serving the same purpose are shown on the same \textit{dashboard}.
We provide a set of intuitive graphs and visual representations of the measured and analyzed metrics, while still allowing users the freedom to modify them or add new metrics according to their needs and preferences.


%

\smallskip{}\section{Implementation}\label{sec:implementation}
The current \sys{} prototype supports the Intel SGX TEE, due to its wide adoption in practice, both in academia and in industry.
Our design is however generic and can be extended for other TEEs.
We implemented the SGX Exporter for the TEE Metrics Exporter component.
We instrumented the Linux SGX driver and developed several eBPF programs to implement the Performance Metrics Exporter component.
Figure~\ref{fig:impl} shows the architecture of \sys{}. 
We rely on Prometheus~\cite{prometheus} (an open-source time-series database) to implement the Performance Metrics Aggregation components, as well as an ad-hoc python program used by the Performance Metrics Analysis (PMAN) to digest raw metrics data and perform threshold analysis.
Lastly, we use Grafana~\cite{grafana}, a widely used framework, to visualize the collected metrics.
We exploit Grafana also for its metric queries, analysis, and various visualization options, thus it's easy to be integrated with the Performance Metrics Analysis component.

The combination of multiple system-wide sources for various metrics allows \sys{} to provide a broad insight into the monitored system and its executed applications. 
In comparison to fine grained performance analysis tools, e.g., \emph{perf}, \sys{} collects key metrics on a system wide level with lower granularity, frequency and only little instrumentation, where perf-like tools may record every event, function call or object change at the application level for advanced analysis features such as call graphs.
\sys{} does not completely trace an application during its runtime but only captures system-wide key events,i.e., important performance metrics, such as an enclave creation or the execution of a system call, thus limiting its impact on overall performance.
Additionally, monitored metrics are usually only counted, requiring very little processing at runtime. 
The data retrieval frequency can be adjusted, by default, the different exporters are queried for data only every 5 seconds. 
In the remainder, we describe additional implementation details and the lessons learned during the process.

\subsection{Performance Metrics Exporter (PME)}

\myparagraph{TEE Metrics Exporter (TME)}
In the implementation of the framework, we decided to initially support the Intel SGX platform, as one of the most mature TEE technologies in the market.
To collect the SGX metrics, we instrument the official Intel SGX driver source code at specific function calls.
Some of the monitored metrics are listed in \S\ref{sec:design}.

All metrics in the SGX driver are instrumented in a similar fashion. 
The use of module parameters (\cite[Chapter 2]{rubini2001linux}) was justified as there is currently no SGX-enabled instrumentation library supporting the C programming language.
Doing so, for each metric, \sloppy{there is a file with the same name in \texttt{/sys/module/isgx/parameters}}.
This is used to write the metric value while the module is loaded and running. 
Our adaptations to the Intel SGX driver span only 42 lines of code.

While we instrument the SGX driver from Intel, \sys only indirectly supports the official Intel SDK for SGX.
\sys{} doesn't monitor OCalls/ECalls specific to the Intel SDK for issuing system calls to the kernel.
However, the system calls resulting from OCalls are indeed monitored by \sys{} at the kernel level.

In order to get the SGX metrics into the aggregation component, an interface component is needed.
It is implemented in Python and takes advantage of the Flask micro-framework~\cite{flask, grinberg2018flask} to expose the metrics via a REST/HTTP endpoint.
The SGX Exporter reads the metrics and exposes them in the OpenMetrics~\cite{openmetrics} format to its metrics endpoint.
From this endpoint, the aggregation component can pull the performance metrics during runtime.

\myparagraph{System Metrics Exporter (SME)}
The SGX driver metrics themselves are insufficient to show all interactions of a trusted application inside enclaves with the untrusted part of the system.
System and infrastructure metrics are additional important cornerstones to monitor applications running inside Intel SGX enclaves.

For instance, the SGX driver only contains the code executed during enclave initialization and functions controlling the EPC pages but it does not interact with other parts of an application's execution cycle such as entering or exiting the enclave, system calls handling, \etc.
These system metrics are provided by three smaller components, each exposing a subset of the overall metrics needed:

\myparagraph{eBPF-Exporter}
Our implementation is based on the exporter for custom eBPF metrics by Cloudflare~\cite{eBPFexporter}.
The exporter contains small eBPF programs written in C to extract metrics during runtime\footnote{\url{https://github.com/iovisor/bcc/}}.
Using kernel probes and trace points, these eBPF programs are executed by the kernel whenever specific events are triggered and enable us to, e.g., count and report occurrences of page faults during runtime.
Metrics currently collected by the eBPF-Exporter consist of \emph{System calls}, \emph{Context switches}, \emph{Page Faults}, and \emph{Cache statistics} while custom eBPF programs can be added if necessary.

\myparagraph{Node-Exporter\footnote{\url{https://github.com/prometheus/node_exporter}}}
The node exporter (written in Go) is part of the Prometheus project and exports machine metrics available through the \emph{/proc} and \emph{/sys} directories in Linux environment.
We integrated the node exporter into \sys{} and reduced the reported metrics to \emph{CPU statistics}, \emph{Memory statistics}, \emph{File system statistics}, and \emph{Network statistics}. 

\myparagraph{cAdvisor\footnote{\url{https://github.com/google/cadvisor}}}
To provide utilization metrics for Docker containers, Google created the cAdvisor web-service. 
We integrated cAdvisor into \sys{} to collect and store per container metrics.
Afterwards, theses metrics are then continuously visualized in a front-end, e.g. Grafana.
The provided metrics can be adjusted depending on the desired level of information and use case~\cite{cadvisordocs}.




\subsection{Performance Metrics Aggregation (PMAG)}
Our monitoring system uses Prometheus~\cite{prometheus} as the Metrics Aggregation Service that collects all metric data from multiple sources and generates insights by querying and aggregating the data.
In particular, we chose Prometheus for its extensibility that allows us to gather internal metrics via a REST/HTTP endpoint.

\subsection{Performance Metrics Visualization (PMV)}
We make use of Grafana to implement the visualization component of \projectname{}.
Grafana is an open-source tool for metrics visualization and analysis~\cite{grafana}. 
We use it to implement \sys{}'s visualization component.
It supports various data sources (\eg{}, Prometheus, InfluxDB, ElasticSearch, \etc{}) and a broad set of visualization widgets (\eg{}, graphs, single stats, tables, heat maps, \etc{}) integrated into the dashboards.


Currently, our prototype consists of three dashboards:
\emph{(i)} an SGX dash\-board showing EPC metrics and a selection of metrics provided by eBPF programs,
\emph{(ii)} a Docker dashboard showing performance data provided by \texttt{cAdvisor} from running Docker containers, and
\emph{(iii)} an infrastructure dashboard showing metrics from both Node-Exporter and eBPF-Exporter.

\begin{figure}[!t]
    \centering
	\includegraphics[width=0.45\textwidth]{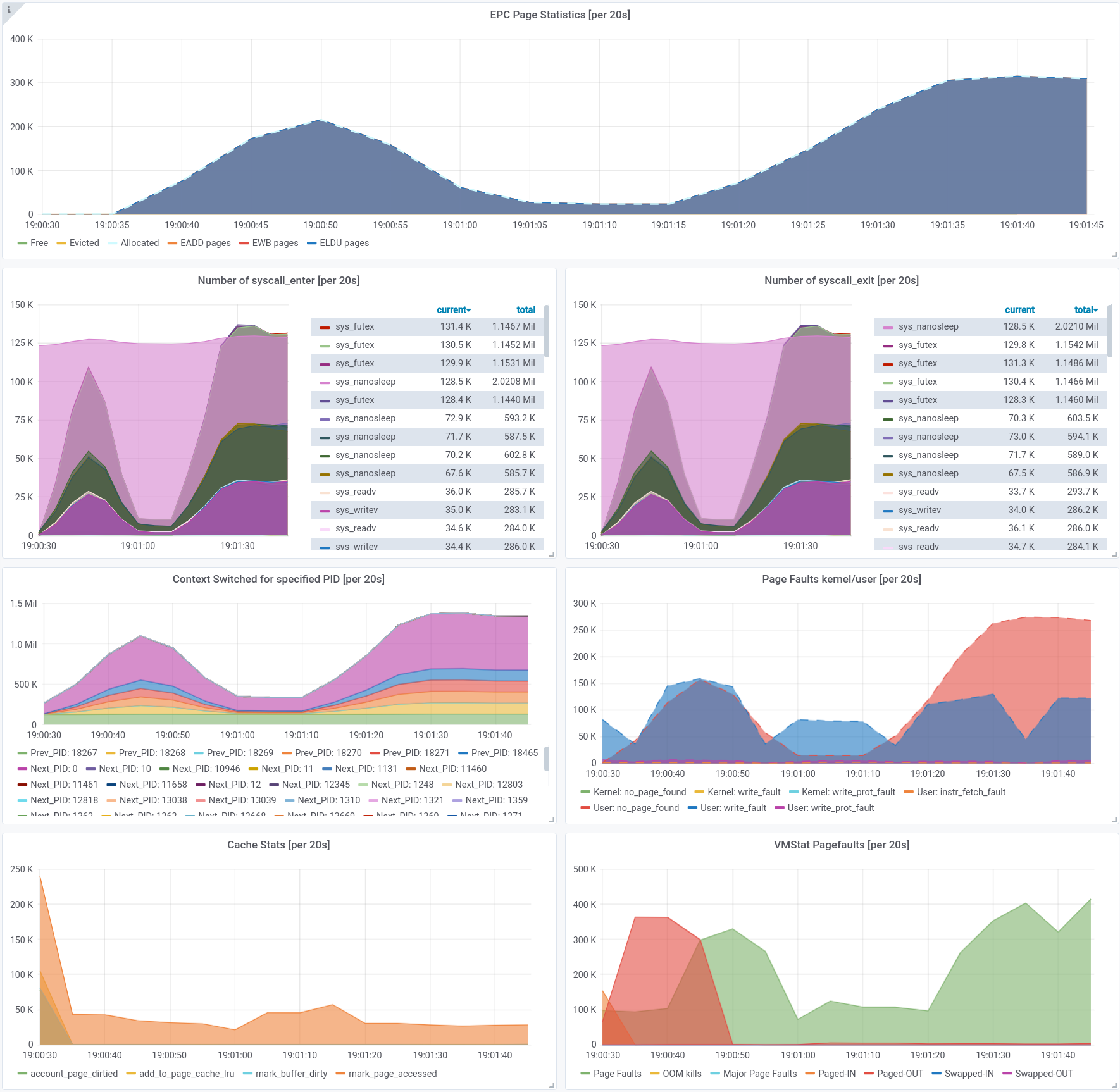}
	\caption{Partial screenshot of \sys{} showing SGX-related metrics.}\label{fig:screenshot}
	\Description[A screenshot of \sys{}]{A screenshot of \sys{}, showing parts of the SGX dashboard in which data of a recorded benchmark is presented as an example.}
\end{figure}

Figure~\ref{fig:screenshot} shows a partial screenshot of the SGX dashboard presented by \sys{}.
Its frontend allows the user to apply a process filter, e.g. \emph{redis-server}, for the continuously updated metrics or select a desired time range from historical data.
The Figure presents recorded data for the Redis database during a benchmark with its two phases (populating the database and executing queries) visible as two consecutive curves.
From the graphs, the user can, for example, study the enclave page cache (top row) utilization, the occurrences of page faults (bottom right), or the distribution of system calls (middle left) during runtime.

\subsection{Deployment with Kubernetes}\label{sec:deployment}
\sys{} components are encapsulated in individual Docker containers for quick deployment on a single host but they can also be deployed in virtual machines or by an orchestrator, such as Kubernetes, an industry-standard for container orchestration.
It features an application-centric design, a well-established API with a uniform set of resources, as well as a powerful ecosystem of third-party tools and extensions.
Its controllers
allow applications and infrastructures to be defined in a declarative manner.
It supports up to \(5000\) server-nodes in a single cluster~\cite{K8large}.

Helm\footnote{\url{https://helm.sh}}, a package manager for Kubernetes applications utilizes \emph{charts} as application definitions for Kubernetes.
These \emph{charts} can easily be deployed, managed, and distributed.
We created a \emph{chart} to install \sys{} in large-scale infrastructures managed by Kubernetes.

In Kubernetes, each of \sys{}'s metrics exporters is deployed (using Helm) in a daemon-like fashion (as \emph{DaemonSet} resource).
\emph{DaemonSets} are deployment configurations that enforce exactly one application instance (pod) running per node in the cluster.
This deployment configuration includes dynamically added nodes. 
Additionally, Kubernetes offers service discovery and resource annotations that \sys{} uses to connect the performance metric aggregation component (\eg{}, Prometheus) to periodically scan for running metrics exporters.
These two features allow \sys{} to adapt to arbitrary changes in the cluster topology.

The \sys{} \emph{chart} also allows for more advanced scheduling scenarios.
For example, in a heterogeneous cluster, 
specific Kubernetes labels, \ie{}, \emph{taints}, can be used to deploy applications based on the availability of hardware features.
Thus, TEE-related metrics exporters can be deployed selectively on nodes that support TEEs, \eg{}, Intel-SGX.
%

\section{Evaluation}\label{sec:evaluation}

This section presents the experimental evaluation of \sys{} using real world applications and various Intel SGX frameworks.
After describing our evaluation settings (\S\ref{sec:hardware-setup}), we evaluate the overhead of \sys{} in monitoring application running inside SGX enclaves, for monitoring overhead (\S\ref{Monitoring_Overheads}) and specific application overheads (\S\ref{Application_Overheads}).
Additionally, we present findings of continuous code profiling in \S\ref{sec:continuous_profiling}.
Next, we show the usability of \sys{} in identifying the cause of performance bottlenecks of state-of-the-art Intel SGX frameworks including SCONE, SGX-LKL, and Graphene-SGX (\S\ref{sgx-framework-comparison}).
Our evaluation demonstrates that the design of \sys{} is generic and it can be transparently used across a variety of SGX frameworks without changing their source code.  

\vspace{5pt}\subsection{Experimental Setup}\label{sec:hardware-setup}

\myparagraph{Testbed}
We used two machines connected via a switched 1~GBit Ethernet network (one hop) to conduct experiments measuring the performance of different Intel SGX frameworks using \sys{}.
The desktop machine is a Fujitsu ESPRIMO P957/E90+ desktop machine with an Intel\textregistered~Core i7\textendash{}6700 CPU, with 32~GB of RAM, running Ubuntu 16.04.6 (kernel v4.15.0) on an SSD SATA-based disk.
The server has an Intel\textregistered~Xeon\textregistered~CPU~E3\textendash{}1280 v6 processor and 64~GB main memory, running Ubuntu 18.04 (kernel v4.15.0\textendash{}70) with the microcode package 3.20191115 installed (microcode revision \texttt{0xca}) to mitigate risks against recent SGX-related attacks~\cite{spectre,meltdown}. 

\myparagraph{Methodology}
For overhead measurements, we report the average evaluation results of 10 runs and examine several configurations for each of the evaluated systems:

\begin{itemize}
	\item Native SGX (without deploying \sys{}), using the official Intel SGX driver. We used the native version as the evaluation baseline.
	\item Activating \sys{} with only TEE Performance Metrics Exporter (PME) component (\ie{}, activating the eBPF-Exporter and the SGX Metrics Exporter).
	\item Activating \sys{} with all components.
\end{itemize}


\vspace{5pt}\subsection{Monitoring Overhead}\label{Monitoring_Overheads}

\begin{figure}[!t]
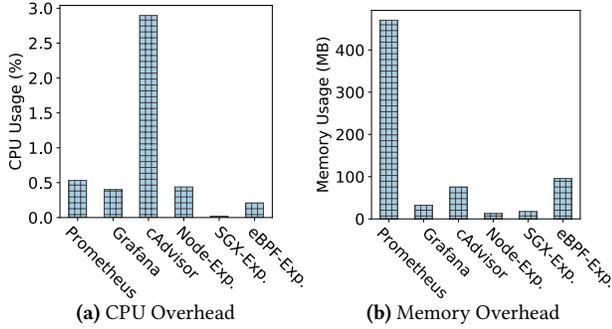

	\captionsetup[subfloat]{farskip=1pt,captionskip=1pt}
	\centering
	\subfloat[CPU Overhead]
	{
	    \includesvg[width=.22\textwidth]{figures/system_monitoring_cpu}\label{subfig:system_monitoring_cpu}
    }
    \subfloat[Memory Overhead]
	{
        \includesvg[width=0.22\textwidth]{figures/system_monitoring_ram}\label{subfig:system_monitoring_ram}
	}
    \caption{The CPU and memory consumption of \projectname's components.}\label{fig:mon-oh2}
    \Description[]{}
\end{figure}





We begin by evaluating the overhead of \sys{} by measuring the CPU and memory footprint of its components, to better understand what is the overhead of each component.
We execute this experiment over 24 hours, by deploying \sys{} on the desktop machine to collect the performance metrics in the system and measure the CPU and Memory usage for each module of each component.


Figure~\ref{fig:mon-oh2} (a) shows the CPU utilization of each component in \sys{}.
Overall, we observe a modest CPU utilization, and at most 3\% on average during the 24~hours for the cAdvisor component.
We envision future versions of \sys{} where this component is deactivated to further reduce interferences induced by the tool itself.

Next, Figure~\ref{fig:mon-oh2} (b) shows the average memory consumption of each monitoring component.
The aggregation component of \sys{} (on top of Prometheus) is the most memory-eager one.
The overall memory footprint of \sys{} is \(\sim\)\(700\)~MB.
While all other components use \(100\)~MB on average, Prometheus allocates 4\(\times\) as much.
This is expected: by design, the aggregator keeps all currently used data in memory for faster retrieval.
Prometheus' memory usage can be reduced by setting a cache limit (default is \(2\)~GB).\footnote{\url{https://prometheus.io/docs/prometheus/1.8/storage/}}

In summary, the overhead of our monitoring framework is negligible for today's standards, allowing users to deploy it in production environments and continuously analyze monitoring data and performance insights of applications running inside SGX enclaves.
As our approach requires no changes to the monitored application and gathers SGX-related statistics at the driver level, no additional memory from the enclave page cache (EPC) is used by \sys{}.

\begin{figure}[!t]
    \centering
    \vspace{3pt} 
    \includesvg[width=.38\textwidth]{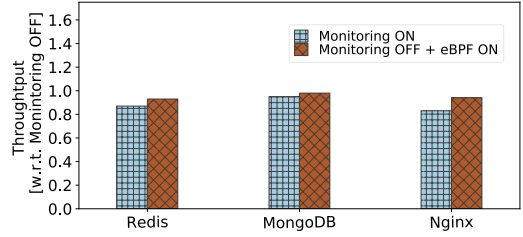}
    \caption{The overhead of \sys{} in monitoring various applications. Results are normalized to the native SGX execution.}\label{fig:mon-oh}
    \Description[]{}
\end{figure}


%
%

\subsection{Overhead with Real-world Applications}\label{Application_Overheads}

We evaluate the overhead induced by \sys{} while monitoring three real-world applications.
Specifically, we use a document database (\ie{}, MongoDB (v3.6.3)~\cite{mongodb}), a web-server (\ie{}, NGINX (v1.14.0)~\cite{nginx}), and an in-memory key-value store (\ie{}, Redis (v5.0.5)~\cite{redis}).
We selected SCONE in this experiment for its ease of use compared to other SGX frameworks.
However, we designed and implemented TEEMon in the way that it can also be used with other SGX frameworks such as SGX-LKL, and Graphene-SGX (see §6.5).

We profile these applications  in three different configurations: \emph{(i)} Native SGX (\texttt{Monitoring OFF}), 
\emph{(ii)} \sys{} with only the PME component (\texttt{Monitoring OFF + eBPF ON}), and \emph{(iii)} full \sys{} with all components (\texttt{Monitoring ON}). 
%
%
Figure~\ref{fig:mon-oh} presents our results. 
We show the average over 10 executions, normalized against the throughput of the native SGX version, without enabling \sys.
The applications' throughput varies from \(87\)\% of baseline executions for NGINX to \(95\)\% for MongoDB.
The eBPF programs running in the kernel contribute for half of the performance drop, and other \sys{} components contribute the other half.
This is expected since several eBPF programs are attached to frequently used performance counters (\eg{}, cache misses, references, system calls, page faults, and context switches).
In some cases (\ie{}, number of context switches) we instrumented both hardware and software counters with \(99\)~Hz and this accounts for some of the added overhead.
This extra overhead can be reduced by disabling unnecessary performance counters, reducing sampling frequency for software counters, or filtering metrics like system calls and context switches to only a specified PID.
To facilitate filtering, we provide a macro for some of the programs which can be set in the eBPF configuration file.



\subsection{Continuous Profiling during code evolution}\label{sec:continuous_profiling}

\begin{figure}[!t]
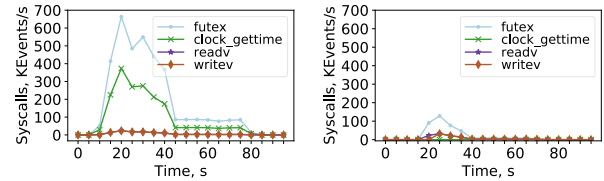

	\captionsetup[subfloat]{farskip=1pt,captionskip=1pt}
	\centering
	\subfloat[Commit \(572bd1a5\)]
	{
	    \includesvg[width=.22\textwidth]{figures/line_plots_before_optimization}\label{subfig:line_plots_before_optimization}
    }
    \subfloat[Commit \(09fea91\)]
	{
        \includesvg[width=0.22\textwidth]{figures/line_plots_after_optimization}\label{subfig:line_plots_after_optimization}
	}
	\caption{Occurrences of selected system calls during execution of Redis with specific versions of \scone{}.}\label{fig:redis-scone-syscalls}
	\Description[]{}
\end{figure}

Besides using it for monitoring deployed applications at a large scale for overall system health and performance, \sys{} can be easily used to track the impact of code changes.
Our approach to continuous integration includes compilation and benchmarks of several applications including Redis while tracking statistics about system calls, page faults, and EPC pages, provided by \sys{}. 

As a showcase, we evaluate two minor releases of \scone{} and run the Redis-benchmark application\footnote{\url{https://redis.io/topics/benchmarks}} using \sys{} to monitor its execution.
This allowed us to find that the \texttt{futex} and \texttt{clock\_gettime} system calls dominated over the read/write system calls to receive and send data. 
Thus, it indicated a performance bottleneck as all system calls trigger an expensive enclave exit.

Figure~\ref{fig:redis-scone-syscalls} shows the occurrences of selected system calls for two consecutive code commits, as monitored by \sys{} during the execution of Redis compiled with \scone{}.
Commit \(572bd1a5\) precedes commit \(09fea91\).
For the specific commit \(572bd1a5\) we found that the \texttt{clock\_gettime} system calls peaked at over \(370\,000\)/sec while read and write system calls were at a tenth of that.

\begin{figure}[!t]
    \vspace{3pt} 
    \centering
    \includesvg[width=.38\textwidth]{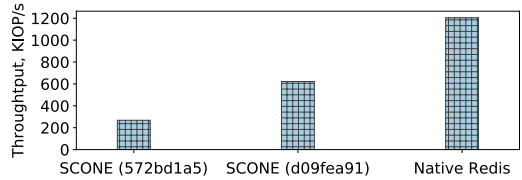}
    \caption{Changes in throughput for Redis at different stages of code evolution and native Redis.}\label{fig:tput_comparison}
    \Description[]{}
\end{figure}

As depicted in Figure~\ref{fig:tput_comparison}, the specific code changes yielded in an almost doubling of average throughput.
The graph shows the performance measurements of Redis using Redis-benchmark on a single host.

With commit \(09fea91\) the handling of the \texttt{clock\_gettime} system call was improved towards handling it on the inside of the enclave without triggering a system call in the kernel.
As a result, the number of \texttt{clock\_gettime} system calls to the kernel and enclave exits is dramatically reduced allowing Redis to handle more requests per second.
With this optimization, we measured at most 100 \texttt{clock\_gettime} system calls per second while the maximum read and write system calls increased from 23 to 32 per second.

For commit \(572bd1a5\), Redis achieves a throughput of \(267\,952.22\)~IOP/s.
With the commit \(09fea91\), the throughput of Redis increased to \(621\,504\)~IOP/s.



\vspace{-5pt}\subsection{Head-to-Head of SGX Framework}\label{sgx-framework-comparison}

We conclude our experimental evaluation by showing that \sys{} is designed in a generic way so that it can be used for different Intel SGX frameworks without changing their source code.
We focus on a head-to-head comparison using the Redis in-memory key-value store as application, deployed and run inside Intel SGX enclaves using several SGX frameworks.
In addition, we also demonstrate that based on the performance metrics and statistics provided by \sys{}, we can identify the cause of bottlenecks of these SGX frameworks.

We benchmarked Redis (v5.0.5) running inside SGX enclaves using SGX-LKL,\footnote{Commit ff8a1a3d, master branch.} SCONE,\footnote{Commit fab5a2b7c, master branch.} and Graphene-SGX.\footnote{Commit e98be31, master branch.}
These SGX frameworks can run legacy applications Intel SGX without changing their code, simply by recompiling or relinking Redis using their provided compilation toolchains.


While Redis was executed directly on the host, we adapted the configuration of Redis to allow for stable execution with all frameworks.
Foremost, we disabled the periodic creation of persistent snapshots, it requires the availability of the \texttt{fork()} system call within the SGX-enclave, which is not available in SGX-LKL and Graphene-SGX.
Furthermore, we configured Redis to use at most \(1\)~GB of memory, \ie{}, the heap size of the enclave configured for all SGX frameworks.

We make use of the \emph{memtier\_benchmark} suite\footnote{\url{https://github.com/RedisLabs/memtier_benchmark}} to measure the performance of Redis and configure it to use \(8\) concurrent threads for optimal performance. 
Hence, the indicated number of connections is always a factor of \(8\).

First, we pre-populate the database with \(720\,000\) keys.
During the measurements, the benchmark issues GET requests.
The \emph{memtier\_benchmark} is configured to use a pipeline of 8 requests and 8 connections per client-thread as these settings provided the best results in preliminary tests.
We run experiments with different Redis database sizes (\(78\)~MB, \(105\)~MB, and \(127\)~MB) by setting the size of values (in the key-value messages) of \(32\)~, \(64\)~, and \(96\)~bytes, respectively.
The reason we conducted the experiments with different database sizes is that the most current SGX hardware supports only \(\sim\)\(94\)~MB EPC size (see \(\S\)\ref{sec:background}) for applications running inside enclaves.
When more memory is required, the applications inside enclaves need to perform the paging mechanism, usually very expensive performance-wise.

Next, we first present the performance comparison of Redis running with different SGX frameworks.
Then, we describe how to use performance metrics captured by \sys{} to identify the bottlenecks of these SGX frameworks.

\begin{figure*}[!ht]
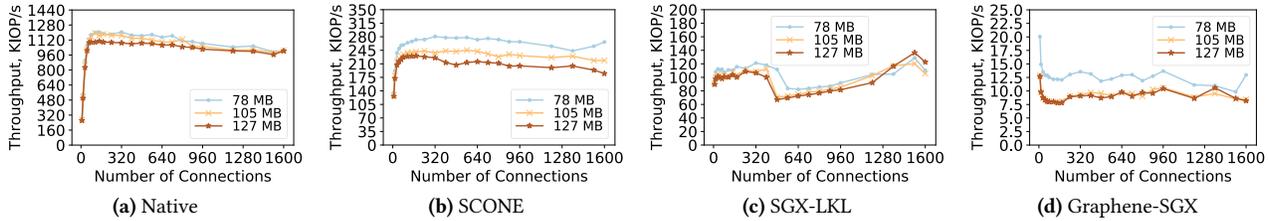

    \setlength{\belowcaptionskip}{-4pt}
    \captionsetup[subfloat]{farskip=2pt,captionskip=1pt}
    \centering
	\subfloat[Native]
	{
	    \includesvg[width=.23\textwidth]{figures/line_plots__Base_all_Throughput}\label{subfig:redis_vanilla_tput}
    }
    \subfloat[SCONE]
	{
        \includesvg[width=.23\textwidth]{figures/line_plots__Scone_all_Throughput}\label{subfig:redis_scone_tput}
    }
    \subfloat[SGX-LKL]
	{
        \includesvg[width=.23\textwidth]{figures/line_plots__LKL_all_Throughput}\label{subfig:redis_lkl_tput}
    }
    \subfloat[Graphene-SGX]
	{
		\includesvg[width=.23\textwidth]{figures/line_plots__Graphene_all_Throughput}\label{subfig:redis_graphene_tput}
	}   
    \caption{The throughput comparison between native Redis and Redis with different SGX frameworks. The total memory usage of Redis is set to different sizes of \(78\)~MB, \(105\)~MB, and \(127\)~MB.
    }\label{fig:redis_tput}\Description[]{}
\end{figure*}

\begin{figure*}[htb!]
    \setlength{\belowcaptionskip}{-4pt}
    \captionsetup[subfloat]{farskip=2pt,captionskip=1pt}
    \centering
	\subfloat[Native]
	{
        \includesvg[width=.23\textwidth]{figures/line_plots__Base_all_Latency}
        \label{subfig:redis_vanilla_lat}
    }
	\subfloat[SCONE]
	{		
        \includesvg[width=.23\textwidth]{figures/line_plots__Scone_all_Latency}
        \label{subfig:redis_scone_lat}
    }
    \subfloat[SGX-LXL]
	{
        \includesvg[width=.23\textwidth]{figures/line_plots__LKL_all_Latency}
        \label{subfig:redis_lkl_lat}
	} 	
	\subfloat[Graphene-SGX]
	{
		\includesvg[width=.23\textwidth]{figures/line_plots__Graphene_all_Latency}
		\label{subfig:redis_graphene_lat}
    }
    \caption{The latency comparison between native Redis and Redis with different SGX frameworks. The total memory usage of Redis is set to different sizes of \(78\)~MB, \(105\)~MB, and \(127\)~MB.
    }\label{fig:redis_lat}
    \Description[]{}
\end{figure*}

\begin{figure}[ht!]
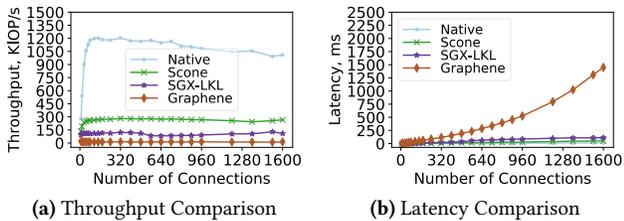

    \captionsetup[subfloat]{farskip=2pt,captionskip=1pt}
    \centering
    \subfloat[Throughput Comparison]
	{
	    \includesvg[width=.23\textwidth]{figures/line_plots__all_Throughput}\label{subfig:redis_all_tput}
    }
    \subfloat[Latency Comparison]
	{
        \includesvg[width=0.23\textwidth]{figures/line_plots__all_Latency}\label{subfig:redis_all_lat}
	}
    \caption{The throughput and latency comparison between native Redis and Redis with different SGX-frameworks. The total memory usage of Redis is set to \(78\)~MB.
    }\label{fig:redis_all_tput_lat}
    \Description[Throughput and Latency measurements]{}
\end{figure}


\begin{figure*}[!ht]
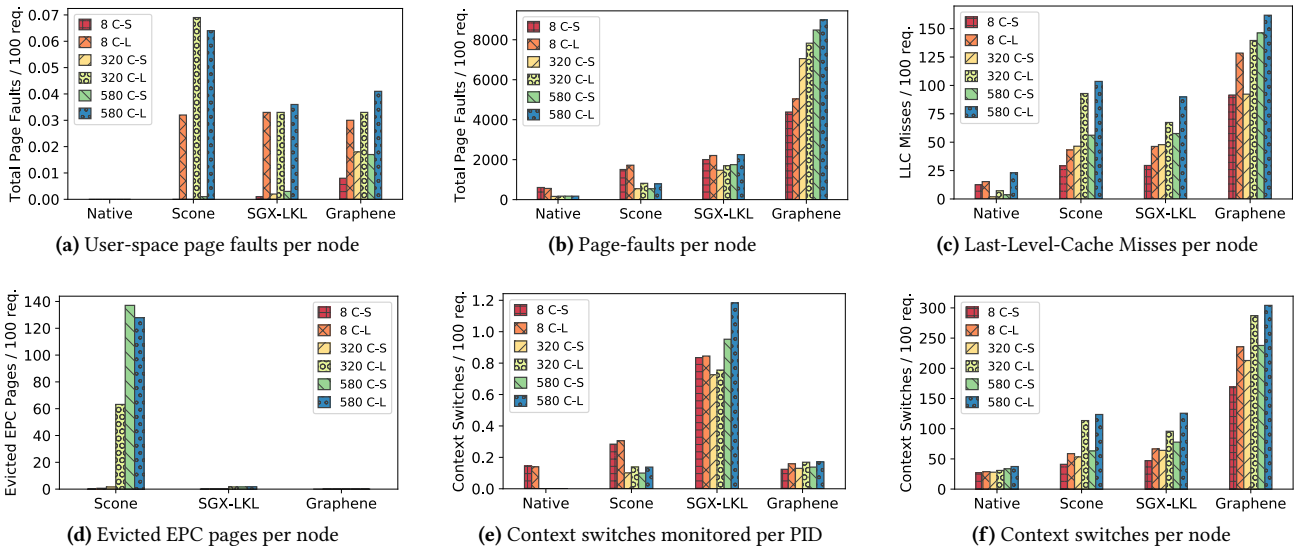

	\captionsetup[subfloat]{farskip=2pt,captionskip=1pt}
	\subfloat[User-space page faults per node]
	{
	\includesvg[width=0.31\textwidth]{figures/ebpf_exporter_page_fault_user_total}\label{fig:ebpf_userpagefaults}
	}
	\hspace{4pt}
	\subfloat[Page-faults per node]
	{
	\includesvg[width=0.31\textwidth]{figures/ebpf_exporter_page_faults_total}\label{fig:ebpf_totalpagefaults}
	}
	\hspace{4pt}	
	\subfloat[Last-Level-Cache Misses per node]
	{
	\includesvg[width=0.31\textwidth]{figures/ebpf_exporter_llc_misses_total}\label{fig:ebpf_llcmisses}
	}

	\vspace{5pt}\hspace{1pt}\subfloat[Evicted EPC pages per node]
	{
	\includesvg[width=0.31\textwidth]{figures/evicted_pages}\label{fig:ebpf_epcpages}
	}
	\hspace{5pt}
	\subfloat[Context switches monitored per PID]
	{
	\includesvg[width=0.31\textwidth]{figures/ebpf_exporter_context_switched_from_pid_total}\label{fig:ebpf_contextsw_pid}
	}
	\hspace{5pt}
	\subfloat[Context switches per node]
	{
	\includesvg[width=0.31\textwidth]{figures/ebpf_exporter_context_switches_total}	\label{fig:ebpf_contextsw_total}
	}
	\caption{The detailed statistics of monitored performance metrics of native Redis and Redis running inside SGX enclaves using different SGX frameworks. The experiments are conducted with different configurations: \(8\)~connections and \(78\)~MB database size (\(8\) C-S); \(8\)~connections and \(105\)~MB database size (\(8\) C-L); \(320\) connections and \(78\)~MB database size (\(320\) C-S); \(320\)~connections and \(105\)~MB database size (\(320\) C-L); \(580\)~connections and \(78\)~MB database size (\(580\) C-S); and for \(580\)~connections and \(105\)~MB database size (\(580\) C-L).} \label{fig:ebpf_plots}
	\Description[]{}
\end{figure*}

\subheading{\#1: Performance Comparison}
In the following, we discuss the performance measurements for Redis running with Intel SGX using SGX-LKL, \scone{}, and Graphene-SGX.
Note that the native version in this experiment is the vanilla Redis running without Intel SGX.

\vspace{2mm}\tightparagraph{Throughput}
Figure~\ref{fig:redis_tput} shows the throughput of Redis using the different SGX frameworks. 
The native Redis achieves the throughput of \(1.01\)~M - \(1.2\)~M input/output operations per second (IOP/s) with different Redis database sizes at \(320\) client connections (Figure~\ref{fig:redis_tput} (a)).
The throughput of native Redis decreases when the number of connections is higher than \(320\).
This is because, above \(320\) client connections, the host's network is squeezed at its capacity of \(1\)~GBps (see \(\S\)\ref{sec:hardware-setup}).

Meanwhile, Figure~\ref{fig:redis_tput}~\subref{subfig:redis_scone_tput} depicts a similar behavior of throughput of Redis running with \scone{}.
The maximum throughput of SCONE is \(278\)~KIOP/s at \(560\) connections (\(\sim\)\(23\%\) throughput of native Redis).
The throughput of Redis with SCONE drops when the database size increases due to the EPC limitation of the SGX hardware.
Increasing the database size from \(87\)~MB to \(105\)~MB reduces the peak performance of Redis with SCONE by \(32\)~KIOP/s (decrease of \(12\)\%).
Further increasing the database size to \(127\)~MB decreases the peak performance at \(29\)~KIOP/s.

Figure~\ref{fig:redis_tput}~\subref{subfig:redis_lkl_tput} shows the results for the throughput of Redis with the SGX-LKL framework.
While Redis with SGX-LKL peaks at \(320\) connections with \(121\)~KIOP/s (\(\sim\)\(10\)\% of native Redis throughput), our results also show a steep drop in performance of Redis with SGX-LKL at \(560\) connections with a steady increase afterward.

Figure~\ref{fig:redis_tput}~\subref{subfig:redis_graphene_tput} shows that, differently from the other SGX frameworks, Graphene-SGX performs best for one client (\(8\) connections) and exhibits a reduced performance for more connections. 
The peak performance of Graphene-SGX was measured at \(20\)~KIOP/s for 8 connections, (\(\sim\)\(1.6\)\% of native Redis throughput).
Similar to SCONE, Figure~\ref{fig:redis_tput}~\subref{subfig:redis_graphene_tput} shows a drop in throughput of Graphene-SGX if the database size increases from \(78\)~MB to \(105\)~MB.
For a single client, the throughput decreases from \(20\)~KIOP/s to \(12\)~KIOP/s.

\vspace{2mm}\tightparagraph{Latency}
Figure~\ref{fig:redis_lat} presents the Redis latency comparison between the different SGX frameworks.
As expected, the latency of all evaluated systems increases when the number of connections increases.
At \(320\) connections, the latency of the native Redis is \(\sim\)\(2\) milliseconds (ms), whereas the latency of Redis with SCONE, SGX-LKL, and Graphene-SGX are \(\sim\)\(9\)~ms, \(\sim\)\(20\)~ms, and \(\sim\)\(249\)~ms, respectively.
All latency measurements show an overall similar correlation between the number of connections and the latency.
However, Redis with Graphene-SGX imposes a significantly higher latency compared to other frameworks.
Figure~\ref{fig:redis_all_tput_lat}~\subref{subfig:redis_all_tput} and \subref{subfig:redis_all_lat} show a performance comparison of native Redis and Redis with different SGX frameworks,
with a database size of \(78\)~MB and an increasing number of clients connections.
In general, these results only show the overall performance trends of the SGX frameworks.
To understand the insights of these SGX frameworks, we analyze the detailed performance metrics data during runtime and identify the bottlenecks using \sys{}.

\subheading{\#2: Performance Metrics Analytics}\balance{}
Figure~\ref{fig:ebpf_plots} presents the performance metrics statistics data of native Redis and Redis with different SGX frameworks, collected during benchmarks as described in \S\ref{sgx-framework-comparison}.1.
All presented statistics and data are similarly presented by the \sys{} front-end during a monitoring session. 

\myparagraph{Page Faults}
Figure~\ref{fig:ebpf_plots}~\subref{fig:ebpf_userpagefaults} and Figure~\ref{fig:ebpf_plots}~\subref{fig:ebpf_totalpagefaults} show the page faults in user space for Redis and the total page faults per host during the benchmark, respectively.
The user space page faults include: \texttt{no\_page\_found}, \texttt{write\_prot\_fault}, \texttt{write\_fault} and \texttt{instr\_fetch\_fault}.

While Figure~\ref{fig:ebpf_plots}~\subref{fig:ebpf_userpagefaults} indicates an overall low rate of user space page faults, it also shows that native Redis causes no page faults in user space.
For the SGX frameworks, the rate of user space page faults increases with database sizes exceeding the EPC size (\(\sim\)\(94\)~MB).
This happens when the SGX-enabled Redis reads data that was previously swapped out of the EPC and is unavailable for the current request.
For \(320\) and \(580\) connections, with the database size of \(105\)~MB, Redis with \scone{} reaches the peaks of \(0.069\) and \(0.064\) user space page faults.
Graphene-SGX and SGX-LKL show a similar pattern of page faults with \(\sim\)\(0.03\) page faults per \(100\) requests for larger database sizes.
While SCONE and SGX-LKL introduce negligible page faults (\eg{}, almost no page fault) with the database size of \(78\)~MB which fits into EPC, the measurements show that Graphene-SGX still has the number of page faults of \(0.02\) per \(100\) requests.

In contrast to the low rate of user level page faults, Figure~\ref{fig:ebpf_plots}~\subref{fig:ebpf_totalpagefaults} shows that on host-wide scope more page faults are registered.
Native Redis has \(607\) total page faults per \(100\) GET requests for \(8\) connections, however, this number decreases (\(\textless\)\(170\) page faults) for larger numbers of connections.
This closely follows the finding that few connections lead to context switches in native Redis.
While SCONE and SGX-LKL have the page fault rates increasing from \(500\) to \(2200\) total page faults per \(100\) GET requests, Graphene-SGX has a significant number of total page faults.
For \(580\) connections and database size of \(105\)~MB, Graphene-SGX has \(8996\) total page faults per \(100\) requests on average.

\myparagraph{Last Level Cache Misses}
Figure~\ref{fig:ebpf_plots}~\subref{fig:ebpf_llcmisses} illustrates the last level cache (LLC) misses during the benchmark.
Compared to native Redis, all SGX frameworks induce an elevated rate of LLC misses.
With native Redis we observe \(1.8-23\) LLC misses per \(100\) GET requests.
Instead, SCONE and SGX-LKL achieve similar (yet higher) rates, \ie{}, \(29\) to \(103\) LLC misses per \(100\) GET requests.
Graphene-SGX has the highest LLC misses: \(91\) for \(8\) connections and \(78\)~MB database size, and up to \(161\) LLC misses for \(580\) connections with \(105\)~MB database size (per \(100\) GET requests).

\myparagraph{Evicted EPC Pages}
Figure~\ref{fig:ebpf_plots}~\subref{fig:ebpf_epcpages} shows the measured evicted pages from the enclave page cache (EPC).
Graphene-SGX has at most \(0.02\) evicted pages per \(100\) GET requests for the database size of \(78\)~MB which fits in the EPC.
For the database size of \(105\)~MB, Graphene-SGX exhibits at most \(0.03\) evicted pages per \(100\) GET requests.
SGX-LKL shows a very similar behavior with up to \(1.6\) evicted pages (per 100 GET requests) for the database size of \(78\)~MB  and up to \(1.7\) evicted pages with the database size of \(105\)~MB.
Meanwhile, SCONE has a stark increase of evicted pages compared to other SGX frameworks.
With the number of connections of \(580\) and for the database size of \(105\)~MB, SCONE has \(137\) evicted pages per \(100\) GET requests.
We attribute the differences to the individual implementation and potential shortcomings of the framework's enclave memory management.

\vspace{2pt}\tightparagraph{Context Switches}
A common cause of SGX performance overheads is costly enclave transitions.
The context switches were filtered by PID, to make it easier to monitor specific applications in the system.


Figure~\ref{fig:ebpf_plots}~\subref{fig:ebpf_contextsw_pid} shows these results and indicates that per \(100\) GET requests, Redis with SGX-LKL hits the most context switches.
Instead, native Redis exhibits 0.14 context switches per 100 requests for the evaluation with just \(8\) connections.
Since Redis uses an event queue and in combination with the findings shown in Figure~\ref{fig:redis_tput}~\subref{subfig:redis_vanilla_tput}, we conclude that, for 8 connections, Redis often waits (sleeps) for new messages and thereby causes context switches.
With the exception of Graphene-SGX, SCONE and SGX-LKL show a similar pattern for 8 connections.

Figure~\ref{fig:ebpf_plots}~\subref{fig:ebpf_contextsw_total} shows the number of total context switches on the host while the GET requests are issued.
The Figure suggests the total (host-wide) context switches of Redis with Gra\-phene-SGX increases dramatically (up to \(12\)\(\times\)) compared to Redis with other SGX frameworks and native Redis.
For 580 connections with a database size of \(105\)~MB, Redis with Graphene-SGX has \(304\) context switches per \(100\) GET requests, while native Redis has only \(37\).
SCONE and SGX-LKL expose a similar pattern as native Redis, with at most \(125\) context switches per \(100\) GET requests.
We believe that Gra\-phene-SGX has the lower performance as shown in Figure~\ref{fig:redis_all_tput_lat} because it has significantly more context switches than the other frameworks as reported by \sys.

Note that Figure~\ref{fig:ebpf_plots}~\subref{fig:ebpf_contextsw_pid} shows only the context switches by Redis process itself, including its threads while Figure~\ref{fig:ebpf_plots}~\subref{fig:ebpf_contextsw_total} shows the total (host-wide) amount of context switches which includes the context switches between kernel processes as well as context switches to the \texttt{ksgxswapd} (Intel® SGX swapping daemon) process.



In summary, in this experiment, we show that \sys{} provides detailed performance data during runtime (e.g., cache misses, context switches, page faults, evicted EPC pages, etc) of applications (e.g., Redis) running inside Intel SGX which helps us to understand the performance behavior of the applications. The presented performance metrics by \sys{} are helpful for developers using SGX frameworks to identify performance issues and to provide guidance for improving the performance of these frameworks, especially with regard to scarce resources such as EPC memory and the expensive enclave exit and enter operations (due to system calls).
This is achieved by presenting valuable graphs that show, e.g., high occurrences of the \emph{clock\_gettime} system call dominating the desired read-write system calls for network IO. 
While different metrics could in principal be gathered individually with different tools, \sys{} provides a single frontend for continuous and effortless monitoring of application to analyse their \emph{behavior} in a production ready environment.

\vspace{-7pt}\section{Conclusion and Future Work}\label{sec:conclusion}

This paper described \sys{}, a real-time performance monitoring framework for applications running inside TEE enclaves.
\sys{} is independent from specific secure execution platforms and applications running on top of it while offering a wide range of performance metrics.
Furthermore, it natively supports a micro-service architecture for applications since all components of \sys{} can be deployed using Docker containers.
We evaluated \sys{} using real-world applications and state-of-the-art SGX frameworks.

Our evaluation shows that \sys{} incurs a very low overhead, which is only from $5$\% to $17$\% depending on running applications,
while it provides valuable insights on the measured performance. 

\sys{} additionally offers a visualization dashboard to inspect in real-time the behavior of the systems being monitored.
The evaluation also shows that \sys{} can be used for many different Intel SGX frameworks without changing the framework's code.
The performance metrics provided by \sys{} allow users to pinpoint bottlenecks and performance issues of applications running inside enclaves using different SGX frameworks and identify the source causing the bottlenecks of these SGX frameworks including SCONE, SGX-LKL and Graphene-SGX.
Finally, \projectname{} has been integrated with Kubernetes to monitor TEE applications running in distributed cluster.

In the future, we will extend and improve \sys{} to offer additional information as well as flexibility to the user.
The probes that \sys{} uses to gather kernel data are currently fixed but in the future, on-demand loading should be possible.
\sys{} will be made available to the research community.

\myparagraph{Acknowledgements}
We thank our shepherd Professor Tim Wood and the anonymous reviewers for their work and helpful comments as well as Rasha Faqeh, Anna Galanou, and F\'abio Silva for their feedback and contribution.
This work was funded by the German Research Foundation, Project-ID 174223256 (TRR 96) and the European Union's Horizon 2020 research and innovation programme under the LEGaTO Project (legato-project.eu), grant agreement No 780681.

{
\bibliographystyle{ACM-Reference-Format}
\bibliography{library}
}
\end{document}